\documentclass[twocolumn,pra,showpacs,superscriptaddress]{revtex4}
\usepackage{amsmath}
\usepackage{graphicx}
\usepackage{amsfonts}

\newcommand{\eq}{\begin{equation}}
\newcommand{\fine}{\end{equation}}

\begin{document}

\title{Quantum to classical transition via fuzzy measurements on
high gain spontaneous parametric down-conversion}
\author{Chiara Vitelli}
\affiliation{Dipartimento di Fisica, ``Sapienza'' Universit\`{a} di Roma,
piazzale Aldo Moro 5, I-00185 Roma, Italy}
\affiliation{Consorzio Nazionale Interuniversitario per le Scienze Fisiche
della Materia, piazzale Aldo Moro 5, I-00185 Roma, Italy}
\author{Nicol\`{o} Spagnolo}
\affiliation{Dipartimento di Fisica, ``Sapienza'' Universit\`{a} di Roma,
piazzale Aldo Moro 5, I-00185 Roma, Italy}
\affiliation{Consorzio Nazionale Interuniversitario per le Scienze Fisiche
della Materia, piazzale Aldo Moro 5, I-00185 Roma, Italy}
\author{Lorenzo Toffoli}
\affiliation{Dipartimento di Fisica, ``Sapienza'' Universit\`{a} di Roma,
piazzale Aldo Moro 5, I-00185 Roma, Italy}
\author{Fabio Sciarrino}
\affiliation{Dipartimento di Fisica, ``Sapienza'' Universit\`{a} di Roma,
piazzale Aldo Moro 5, I-00185 Roma, Italy}
\affiliation{Istituto Nazionale di Ottica Applicata, largo Fermi 6, I-50125 Firenze, Italy}
\author{Francesco De Martini}
\affiliation{Dipartimento di Fisica, ``Sapienza'' Universit\`{a} di Roma,
piazzale Aldo Moro 5, I-00185 Roma, Italy}
\affiliation{Accademia Nazionale dei Lincei, via della Lungara 10, I-00165 Roma,
Italy}

\begin{abstract}
We consider the high gain spontaneous parametric down-conversion
in a non collinear geometry as a paradigmatic scenario to
investigate the quantum-to-classical transition by increasing the
pump power, that is, the average number of generated photons. The
possibility of observing quantum correlations in such macroscopic
quantum system through dichotomic measurement will be analyzed by
addressing two different measurement schemes, based on different
dichotomization processes. More specifically, we will investigate
the persistence of non-locality in an increasing size
$\frac{n}{2}$-spin singlet state by studying the change in the
correlations form as $n$ increases, both in the ideal case and in
presence of losses. We observe a fast decrease in the amount of
Bell's inequality  violation for increasing system size.  This
theoretical analysis is supported by the experimental observation
of macro-macro correlations with an average number of photons of
about $10^{3}$. Our results enlighten the practical extreme difficulty of
observing non-locality by performing such a dichotomic fuzzy
measurement.
\end{abstract}

\maketitle

\section{Introduction}

For long time the investigation about entanglement and
non-locality has been limited to quantum systems of small size
\cite{Bell64}. Theoretical and experimental works on Bell's
inequalities have been devoted to the study of single particle
states, in which dichotomic measurements have been performed
\cite{Clau69}. Non-locality tests have been achieved with single
photon states, produced by parametric down conversion, by
detecting polarization correlations \cite{Alle90,Ou88,Kies98}.
More recently the violation of Bell's inequality has been
performed with a larger number of photons: on GHZ  \cite{Chen06}
and cluster states \cite{Walt05} up to 4 photons.\\
On the other hand, the possibility of observing quantum phenomena
at a macroscopic level seems to be in conflict with the classical
description of our everyday world knowledge. The main problem for
such observation arises from the experimental difficulty of
sufficiently isolating a quantum system from its environment,
i.e., from the decoherence process \cite{Zure03}.
An alterative approach to explain the quantum-to-classical transition,
conceptually different from the decoherence
program, has been given, very recently, by Kofler and Brukner,
along the idea earlier discussed by Bell, Peres \cite{Pere93} and others.
They have given a description of the emergence of macroscopic
realism and classical physics in systems of increasing size
\textit{within quantum theory} \cite{Kofl07}.
They focused on the limits of the quantum effects
observability in macroscopic objects, showing that, for large
systems, macrorealism arises under coarse-grained measurements.
More specifically they demonstrated that, while the evolution of a
large spin cannot be described classically when sharp measurement
are performed, a fuzzy measurement on a large spin system would
induce the emergence of the Newtonian time evolution from a full
quantum description of the spin state. However, some counterexamples
to such modelization have been found later by the same authors: some non
classical Hamiltonians violate macrorealism despite coarse-grained
measurements \cite{Kofl08}. One example is given by the
time-dependent Schr\"{o}dinger catlike superposition, which can
violate macrorealism by adopting a suitable ``which emisphere''
measurement. Therefore the measurement problem seems to be a key
ingredient in the attempt of understanding the limits of the
quantum behavior of physical systems and the quantum-to-classical
transition question. As a further step, Kofler, Bruic and Brukner also demonstrated \cite{Kofl09}
that macrorealism does not imply a continuous spatiotemporal
evolution. Indeed, they showed that the same Schr\"{o}dinger catlike
non-classical Hamiltonian, in contact with a dephasing environment
does not violate any longer a Leggett-Garg inequality, while it still
presents a non-classical time evolution. In a recent paper Jeong et al.
\cite{Jeon09} contributed to the investigation about the possibility of observing the quantum features of a system when fuzzy measurement are performed on it, finding that extremely-coarse-grained measurements can still be useful to reveal the quantum world where local realism fails.

In this context, the possibility of obtaining macroscopic quantum
systems in laboratory has raised the problem of investigating
entanglement and non locality in systems in which single particles
cannot be addressed singularly. As shown in Ref. \cite{Chen02},
the demonstration of non-locality in a multiphoton state produced
by a non-degenerate optical parametric amplifier would require the
experimental application of parity operators. On the other hand,
the estimation of a coarse grained quantity, through collective
measurements as the ones proposed in Ref. \cite{Port06}, would
miss the underlying quantum structure of the generated state,
introducing elements of local realism even in presence of strong
entanglement and in absence of decoherence. The theoretical
investigation on a multiphoton  system, obtained via parametric
down conversion, has been also carried out  by Reid et al.
\cite{Reid02}. They analyzed the possibility of obtaining the
violation of Bell's inequality by performing dichotomic
measurement on the multiparticle quantum state.
More specifically, in analogy with the spin formalism, they
proposed to compare the number of photons polarized ``up'' with
the number of photons polarized ``down'' at the exit of the
amplifier. The result of this comparison could be either (+1) or
(-1) hence the measurement on the multiphoton state turned out to be
dichotomic. In such a way Reid et al. revealed a small violation
of the multiparticle Bell's inequality even in presence of losses
and quantum inefficiency of detectors. It's worth noting that
this violation presents a fast decreasing behavior as a function
of the generated photons number. In a recent paper, Bancal et al.
\cite{Banc08} have discussed different techniques for testing
Bell's inequalities in multipair scenarios, in which either at
Alice's and Bob's site a global measurement is performed. They
distinguished between two cases: distinguishable, i.e.
independent, and indistinguishable, i.e. belonging to the same
spatial and temporal mode, photon pairs. They found that while the
state of indistinguishable pairs results more entangled, the state
of independent pairs appears to be more nonlocal.

In the present manuscript, we investigate the
macroscopic-macroscopic state generated by high gain spontaneous
parametric down-conversion. The possibility of observing quantum
correlations in macroscopic quantum systems through dichotomic
measurement will be analyzed, by addressing two different
measurement schemes, based on different dichotomization processes.
More specifically, we will investigate the persistence of
non-locality in an increasing size $\frac{n}{2}$-spin singlet
state by studying the change in the correlations form as $n$
increases, both in the ideal case and in presence of losses. At
last, experimental observation of macro-macro correlations will be
reported. The results obtained enlighten that dichotomic
fuzzy measurements lack of the necessary resolution to characterize
such states and show the extreme difficulty to observe quantum
non-locality in this experimental configuration.

Let us give a brief outline of the paper. Section \ref{sec:Macro_Macro_system} is devoted
to the introduction of the analyzed quantum system. In Section
\ref{sec:Dichotomic_meas}, we describe two different types of
dichotomic measurements on multiphoton states: the orthogonality
filtering (OF) and the threshold detection (TD). In Section
\ref{sec:Bell's_inequality} we present the results of the
numerical simulations of the correlations between dichotomic
measurements carried out on the multiphoton fields produced via
spontaneous parametric down conversion. We report the theoretical
interference fringe-patterns for single-$\frac{n}{2}$ states, we
observe a transition from the sinusoidal pattern of the
spin-$\frac{1}{2}$ into a quasi-linear pattern by increasing the
number of photons of the spin state. According to this behavior we
observe a progressive decrease in the amount of the violation as
also predicted in Ref \cite{Reid02,Banc08}. As following step, we
estimate the correlation after propagation over a lossy channel.
We then analyze the response of the system to the two dichotomic
measurements reported in Sec.\ref{sec:Dichotomic_meas}, discussing
the feasibility of a CHSH test with these detection strategies.
Finally, Section \ref{sec:experimental} is dedicated to an
experimental test of the previous results.

\section{Macroscopic quantum state based on high gain spontaneous parametric
down-conversion}\label{sec:Macro_Macro_system} 
\begin{figure}[th]
\centering
\includegraphics[width=0.5\textwidth]{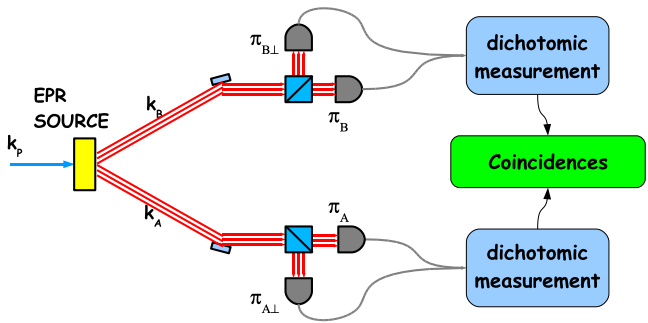}
\caption{(Color online) Scheme of the macroscopic-macroscopic source
and of the detection apparatus. The multiphoton fields on the two
spatial modes of an EPR source are analyzed in polarization with
dichotomic measurements, where the $+1$ value is assigned if
$n_{\protect\pi }>n_{\protect\pi _{\bot }}$, and $-1$ otherwise.
Finally, coincidences between the two apparata are considered.}
\label{fig:conceptual_scheme_spont}
\end{figure}

The investigation
on the micro-macro transition will be performed on a paradigmatic
physical system: the optical parametric amplifier working in a
high gain regime. The quantum state produced in a low gain regime
has been experimentally realized and deeply studied in the past
few years \cite{Eise04,Cami06}. We are now interested in analyzing
the behavior of such quantum system when the number of photons is increased and
it undergoes a fuzzy measurement, in which the generated particles
cannot be addressed singularly, but a dichotomic measurement is
performed on the overall state. More specifically, the radiation
field under investigation is the quantum state obtained by
spontaneous parametric down-conversion (SPDC) with an EPR type-II
source \cite{Kwia95,Eise04}, whose interaction Hamiltonian
is: $\mathcal{H}_{int} = \imath \hbar \chi \left( \hat{a}^{\dag}_{\pi} \hat{b%
}^{\dag}_{\pi_{\bot}} - \hat{a}^{\dag}_{\pi_{\bot}}
\hat{b}^{\dag}_{\pi}
\right) + \mathrm{H.c.}$ where $\hat{a}^{\dag}_{\pi}$ and $\hat{b%
}^{\dag}_{\pi} $\ are the creation operators corresponding to the
generation of a $\pi$-polarized photon on spatial modes
$\mathbf{k}_{A}$ and $\mathbf{k}_{B}$, as sketched in Fig.\ref
{fig:conceptual_scheme_spont}, and $\chi$ is the constant describing
the strength of the interaction. The adoption of the single mode interaction Hamiltonian $\mathcal{H}_{int}$ to model 
the SPDC source is valid since in our experimental setup, described in Sec.\ref{sec:experimental}, 
spatial and spectral filtering are performed on the generated field. A detailed analysis 
of the model taking into account a broadband  pump pulse can be found in reference \cite{Gric97}. 

\noindent The output state reads \cite{Simo03,Eise04,Cami06}:
\begin{equation}  \label{eq:SPDC_state}
|\Psi^{-}\rangle = \frac{1}{C^{2}} \sum_{n=0}^{\infty} \Gamma^{n}
\sqrt{n+1} |\vert \psi^{-}_{n}\rangle
\end{equation}
{\small
\begin{equation}  \label{eq:singlet_n}
|\psi^{-}_{n}\rangle = \frac{1}{\sqrt{n+1}} \sum_{m=0}^{n}
(-1)^{m} |(n-m)\pi,m \pi_{\bot}\rangle_{A} |m
\pi,(n-m)\pi_{\bot}\rangle_{B}
\end{equation}
}where $\Gamma = \tanh g$ and $C = \cosh g$; $g = \chi t$ is the
non-linear gain (NL) of the process. Hence, the output state can
be written as the
weighted coherent superposition of singlet spin-$\frac{n}{2}$ states $%
|\psi^{-}_{n}\rangle$.

As said, this EPR source has been already studied in different
gain regimes. First, Kwiat et al. \cite{Kwia95} exploited the
polarization singlet-state emitted in the single-pair regime to
obtain the violation of Bell's inequalities. Subsequent works
studied the multi-photon states generated in a high gain SPDC
source. Eibl. et al. \cite{Eibl03} experimentally demonstrated
four-photon entanglement in the second-order emission state of the
SPDC source, by taking the four-fold coincidences after the two
output modes of the source were splitted by 50-50 beam-splitters.
A generalized non-locality test \cite{Wein01} was also successfully
performed in this configuration. A similar scheme was subsequently
exploited by Wieczorek et al. \cite{Wiec08} to experimentally generate
an entire family of four-photon entangled states.
The presence of polarization-entanglement in the
multi-photon states up to 12 photons has been proved by studying
the high losses regime where at most one photon per branch was
detected \cite{Eise04}. The density matrix of this two-photons
state was analytically derived and experimentally investigated in
a more recent work \cite{Cami06}, where it has been demonstrated
that it coincides with the one of a Werner state (WS), i.e., a
weighted superposition of a maximally entangled singlet state with
a fully mixed state. Simon and Bouwmeester \cite{Simo03} derived a criteria to
quantify the entanglement of the multiphoton states produced by a high gain SPDC 
source by measuring the Stokes parameters of  polarization of the beams A and B. 
For a high efficiency detection observation of entanglement was predicted. 
However, no experimental demonstration of
entanglement and non-locality has been given in the multi-photon
regime where the generated state does not undergo to a controlled
lossy detection scheme.

\section{Dichotomic measurements on macroscopic states}

\label{sec:Dichotomic_meas}In the context of the investigation on
entanglement and non-locality between macroscopic systems, Bell's
inequalities have been generalized to many particle regimes. Among
various strategies, several possible extensions of dichotomic
measurements in the macroscopic regime have been presented
\cite{Reid02,Banc08}. By these methods, CHSH-type inequalities can
be exploited in order to perform non-locality tests also in
many-particle collective states.

In this section we analyze two possible kind of dichotomic
measurements on macroscopic states, based on photon counting and
signal processing techniques. The first technique is based on the
Orthogonality-Filter (O-Filter) device \cite{Naga07,DeMa08}, which
has already been used to test experimentally the entanglement
between a microscopic and a macroscopic field. This method
has several analogies with the detection strategy presented in Ref. \cite{Seka09}
and attributed to a biological "human eye" detector.
The second technique is a threshold dichotomic
detection scheme, whose action is independent on the input state
and hence can be fairly exploited for a Bell's inequalities test.

\begin{figure}[ht]
\centering
\includegraphics[width=0.4\textwidth]{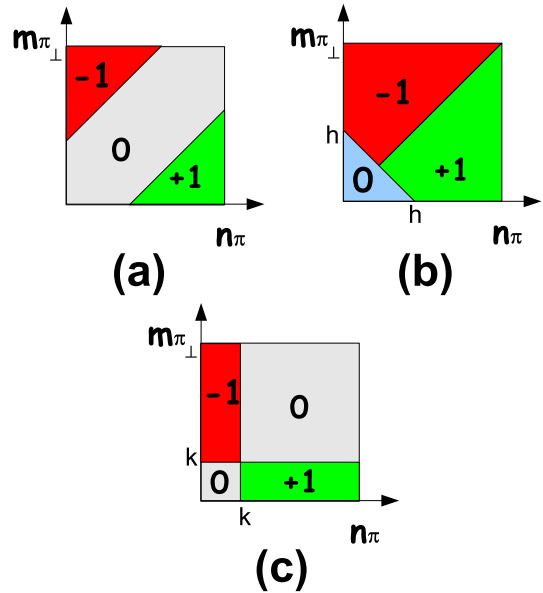}
\caption{(Color online) Selected regions of the Fock space for the two presented
measurement schemes. Each diagram in this figure is referred to
a single spatial mode.
(a) O-Filtering technique representation in the bidimensional
Fock-Space $\left\{ n_{\protect\pi}, m_{\protect\pi_{\bot}} \right\}$. The
(+1) and (-1) regions correspond to a difference in the detected photon
numbers $\vert n_{\protect\pi} - m_{\protect\pi_{\bot}} \vert > k$. The (0)
region corresponds to an inconclusive measurement. (b) Dichotomic threshold
measurement representation in the bidimensional Fock-Space $\left\{ n_{%
\protect\pi}, m_{\protect\pi_{\bot}} \right\}$. Only those pulses
containing a sufficiently high photon number can be detected due
to the threshold response of the apparatus. Then, a dichotomic
assignment is performed on the measurement outcomes. (c) Response of
``biological'' detectors, which are sensible to the impinging
field only if the photon number exceeds an intrinsic threshold $k$.}
\label{fig:measurement_schemes}
\end{figure}

\subsection{Orthogonality Filtering}
\label{subsec:o-filtering_detection}

The first dichotomic measurement technique we analyze in this section is
based on the O-Filter (OF) device introduced in \cite{Naga07,DeMa08}.
In these papers the
detection method allowed to discriminate among two macroscopic orthogonal states $
\left\{ |\Phi ^{\phi}\rangle ,|\Phi ^{\phi_{\bot}}\rangle \right\}
$, obtained by the collinear parametric amplification of
single-photon states with equatorial polarization $\left\{
\vec{\pi}_{\phi},\vec{\pi}_{\phi_{\bot}}\right\}$, defined as
$\vec{\pi}_{\phi} = 2^{-1/2} \left( \vec{\pi}_{H} +
e^{\imath \phi} \vec{\pi}_{V} \right)$, by exploiting the
difference in their photon number distributions. We utilize now this
technique in a different experimental framework. Let us now give a formal
description of this measurement technique in the POVM framework.
First, the incident radiation is analyzed in polarization
by a couple of photon-number resolving detectors
on each spatial mode $\left\{ \mathbf{k}_{A}, \mathbf{k}_{B} \right\}$.
In the ideal case, this measurement corresponds to the projection
of the impinging field onto the Von Neumann operators: $\hat{\Pi}%
_{n,m}=|n\pi ,m\pi _{\bot }\rangle \,\langle n\pi ,m\pi _{\bot }|$, where $%
|n\pi ,m\pi _{\bot }\rangle $ represents a quantum state with $n$
photons with polarization $\pi $ and $m$ photons with polarization
$\pi _{\bot }$. Subsequently, the dichotomization of the
measurement corresponds to assign the value (+1) if $n_{\pi
}-m_{\pi _{\bot }}>k$, (-1) if $m_{\pi _{\bot }}-n_{\pi }>k$, and
(0) otherwise (Fig.\ref {fig:measurement_schemes}-(a)). This
choice of the detection scheme corresponds to the POVM operators:
\begin{eqnarray}
\label{eq:O-Filtering_POVM_1}
\hat{F}_{\pi ,\pi _{\bot }}^{(+1)}(k) &=&\sum_{n=k}^{\infty }\sum_{m=0}^{n-k}%
\hat{\Pi}_{n,m} \\
\hat{F}_{\pi ,\pi _{\bot }}^{(-1)}(k) &=&\sum_{m=k}^{\infty }\sum_{n=0}^{m-k}%
\hat{\Pi}_{n,m} \\
\label{eq:O-Filtering_POVM_3}
\hat{F}_{\pi ,\pi _{\bot }}^{(0)}(k) &=&\hat{I}-\hat{F}_{\pi ,\pi _{\bot }}^{(+1)}-\hat{F}%
_{\pi ,\pi _{\bot }}^{(-1)}
\end{eqnarray}
The discarded outcome ( gray (0) region in Fig.\ref{fig:measurement_schemes}-(a))
turns out to be state dependent. This property, as we shall see later,
renders this kind of dichotomic measurement not fair for applications
in Bell's inequalities test.

\subsection{Threshold detection}

\label{subsec:threshold_detection}

We now introduce a different dichotomic measurement method which
is based on a threshold detection scheme. Let us consider the
following apparatus. As in the OF case, the incident field is
analyzed in polarization on each spatial mode by photon-counting
detectors, and the Von Neumann operators that describe this
intensity measurement are again the $\hat{\Pi}_{n,m}$ projectors.
The dichotomization of the measurement then proceeds as follows (Fig.
\ref{fig:measurement_schemes}-(b)). The (+1) outcome is assigned
when the threshold condition $n_{\pi}+m_{\pi_{\bot}}>h$ is satisfied and when
$n_{\pi}>m_{\pi_{\bot}}$.
Analogously, the (-1) outcome is assigned in the opposite case
$n_{\pi}<m_{\pi_{\bot}}$ conditionally to the threshold
condition: $n_{\pi}+m_{\pi_{\bot}}>h$. If $n_{\pi} = m_{\pi_{\bot}}$, one of the two outputs
($\pm 1$) is randomly assigned with equal probability $p=1/2$.
The POVM operators that describe the measurement can then be
written in the form:
\begin{eqnarray}
\label{eq:threshold_POVM_1}
\hat{T}_{\pi ,\pi _{\bot }}^{(+1)}(h) &=&\sum_{n=h}^{\infty }\sum_{m<\frac{n%
}{2}}\hat{\Pi}_{n-m,m}\\
\hat{T}_{\pi ,\pi _{\bot }}^{(-1)}(h) &=&\sum_{n=h}^{\infty }\sum_{m>\frac{n%
}{2}}\hat{\Pi}_{n-m,m} \\
\label{eq:threshold_POVM_3}
\hat{T}_{\pi ,\pi _{\bot }}^{(0)}(h) &=&\hat{I}-\hat{T}_{\pi ,\pi _{\bot }}^{(+1)}-\hat{T}%
_{\pi ,\pi _{\bot }}^{(-1)}
\end{eqnarray}
This scheme has the peculiar property of selecting an invariant region
of the Fock space with respect to rotations of the polarization basis.
More specifically, let us consider the case in which the measurement is
performed choosing a polarization basis $\pi, \pi_{\bot}$. With that choice, all
the pulses for which $n_{\pi} + m_{\pi_{\bot}} \leq h$ are not detected.
Rotating the basis to $\pi^{'}, \pi_{\bot}^{'}$, the undetected part of the wave function
still corresponds to the application of the same threshold condition in the
new basis $n_{\pi^{'}} + m_{\pi_{\bot}}^{'} > h$. Hence, the filtered Fock-space
region is independent on the choice of the polarization basis but is a function only
of the threshold $h$, which is an intrinsic property of the detection apparatus.
This feature is the main difference with the OF device discussed in previous
section, and renders the TD based detection strategy feasible for its implementation
in Bell's inequalities tests.

\section{Bell's inequalities between macroscopic photonic states generated by
high gain spontaneous-parametric down-conversion}
\label{sec:Bell's_inequality}

In this section we perform a theoretical investigation on
non-locality in a specific macroscopic quantum system analyzed
with the threshold detection apparatus previously introduced.
Specifically, we investigate the quantum correlations between the
fields associated to modes $\mathbf{k}_{A}$ and $\mathbf{k}_{B}$
of the state of Eq.(\ref{eq:SPDC_state}) by using the dichotomic
measurements described in the previous section. More specifically,
we derive the interference fringe pattern obtained by varying the
polarization analysis basis on mode $\mathbf{k}_{B}$. As a first
step, we consider individually the singlet spin-$\frac{n}{2}$
states (\ref {eq:singlet_n}), and we subsequently extend the
results to the SPDC output superposition state of Eq.
(\ref{eq:SPDC_state}). To generalize the results to a realistic
detection and transmission apparatus, we numerically simulate the
effect of losses and non-unitary detection efficiency in the
interference fringe pattern. Finally, we address the problem of
non-locality by investigating a CHSH inequality for this detection
apparatus, studying how the amount of violation is modified by the
increase in the photon's number $n$.

\subsection{Interference fringe pattern on singlet spin-$\frac{n}{2}$ states}

We begin our analysis on the macroscopic-macroscopic state by evaluating
the correlations existing between the two spatial modes of the spin-$%
\frac{n}{2}$ singlet states (Eq.(\ref{eq:singlet_n})). We use a pure
dichotomic measurement scheme, where the (+1) and (-1) outcomes
are assigned whether the difference in the number of photons with
two orthogonal polarization is positive or negative. Finally, if the detected difference
in the number of photons is $0$, one of the ($\pm1$) outcomes is randomly
assigned to the event with equal probability $p=1/2$. We note that
this choice is a subcase of the threshold detection and
O-filtering methods introduced in the previous Section,
corresponding to the values $h=0$ and $k=0$.

\begin{figure}[ht]
\centering
\includegraphics[width=0.5\textwidth]{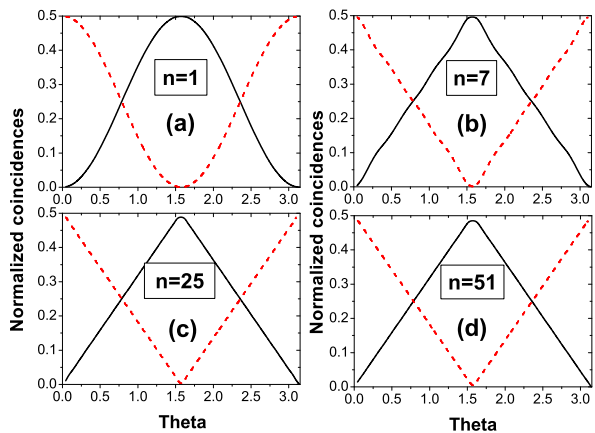}
\caption{(Color online) Theoretical interference fringe-patterns for singlet spin-$\frac{n}{%
2}$ states. The polarization basis on mode $\mathbf{k}_{A}$ is kept fixed
while on mode $\mathbf{k}_{B}$ the basis is varied to obtain the fringe
pattern. Figures correspond to values of (a) $n=1$, (b) $n=7$, (c) $n=25$
and (d) $n=51$. The sinusoidal pattern of the spin-$\frac{1}{2}$ progressively
transforms into a linear pattern. In all figures, black continuous line corresponds  to
the coincidences of both the (+1,+1) and (-1,-1)  outcome configurations,
while red dashed line corresponds to the (+1,-1) and (-1,+1) outcomes on the
two spatial mode. Note that the maximum for each fringe is $0.5$, which is the
probability to obtain one of the two possible anti-correlated outcomes ($\mp 1, \pm 1$).}
\label{fig:fringe_pattern_fixed_n_spont}
\end{figure}

The scheme for evaluating the correlations is sketched in Fig. \ref
{fig:conceptual_scheme_spont}. The two spatial modes of the $|\psi
_{n}^{-}\rangle $ are analyzed with the dichotomic measurement apparatus
here described. The polarization's basis on mode $\mathbf{k}_{A}$ is fixed on $%
\left\{ \vec{\pi}_{+},\vec{\pi}_{-}\right\} $, while on mode
$\mathbf{k}_{B}$ the analysis  basis is varied over the Bloch
sphere. In particular, due to the SU(2) symmetry of the emitted states,
it is sufficient to consider only
the linear polarizations case, defined by the rotation: $\vec{\pi}%
_{\theta }=\cos \theta \,\vec{\pi}_{+}+\sin \theta
\,\vec{\pi}_{-}$. The fringe patterns are then obtained by
evaluating the coincidences between the
outcomes of the two detection apparatus on modes $\mathbf{k}_{A}$ and $%
\mathbf{k}_{B}$. More specifically, this measurement corresponds to the
evaluation of the averages:
\begin{equation}
\begin{aligned} \label{eq:fringe_def_dichotomic} D^{(\pm 1,\pm 1)}_{\vert
\psi^{-}_{n} \rangle}(\theta) &= \langle \psi^{-}_{n} \vert \left(
\hat{T}^{(\pm 1)}_{+,-}(0) \right)_{A} \otimes \left( \hat{T}^{(\pm
1)}_{\theta, \theta_{\bot}}(0) \right)_{B} \vert \psi^{-}_{n} \rangle \\ &=
\langle \psi^{-}_{n} \vert \left( \hat{F}^{(\pm 1)}_{+,-}(0) \right)_{A}
\otimes \left( \hat{F}^{(\pm 1)}_{\theta, \theta_{\bot}}(0) \right)_{B}
\vert \psi^{-}_{n} \rangle \end{aligned}
\end{equation}
The calculation of this measurement has been analytically performed by
expressing the singlet spin-$\frac{n}{2}$ states of eq.(\ref{eq:singlet_n})
in the analyzed polarization basis:
\begin{equation}
|\psi _{n}^{-}\rangle =\sum_{m=0}^{n}\sum_{p=0}^{n}\epsilon
_{m,p}^{n}(\theta )\left| (n-m)+,m-\right\rangle _{A}\left| p\theta
,(n-p)\theta _{\bot }\right\rangle _{B}
\end{equation}
where:
\begin{equation}
\label{eq:coeffic_rotated}
\begin{aligned} \epsilon_{m,p}^{n}(\theta) &= \sum_{q(m,p)} (-1)^{q}
\alpha_{\theta}^{m+p-2q} \beta_{\theta}^{n-m-p+2q} \\ &\left[ C^{n-m}_{p-q}
C^{n-p}_{m-q} C^{m}_{q} C^{p}_{q} \right]^{\frac{1}{2}} \end{aligned}
\end{equation}
with $\alpha _{\theta }=\cos \theta $, $\beta _{\theta }=\sin \theta $ and $%
C_{j}^{i}=\frac{i!}{j!(i-j)!}$ is the binomial coefficient. The
limits of the sum over $q$ have an explicit dependence on the
values of $p$ and $m$ and are not reported here. Finally, by direct
application of the measurement operator, the interference fringe
patterns are evaluated as:
\begin{equation}
\label{eq:int_fringe_pattern_n}
D_{|\psi _{n}^{-}\rangle }^{(\pm 1,\pm 1)}(\theta
)=\sum_{\{m,p\}}|\epsilon _{m,p}^{n}(\theta)|^{2}
\end{equation}
The extension of the sums over $m$ and $p$ depends on the choice of the
outcome on each spatial mode according to the definitions of Eqs.(\ref
{eq:O-Filtering_POVM_1}-\ref{eq:O-Filtering_POVM_3}) and (\ref
{eq:threshold_POVM_1}-\ref{eq:threshold_POVM_3}).

In Fig.\ref{fig:fringe_pattern_fixed_n_spont} we report the
results obtained for different values of the number of photons
$n$. The simplest case, corresponding to a spin-$\frac{1}{2}$
state, presents the well-known sinusoidal pattern, as shown in
Fig. \ref{fig:fringe_pattern_fixed_n_spont}-(a). The sinusoidal
pattern is responsible for the violation of Bell's inequalities as
no classical system can present this dependence from the
phase $\theta $. For progressively higher values of $n$, as shown
in Fig. \ref{fig:fringe_pattern_fixed_n_spont}-(b-d), the fringe
pattern changes its dependence from the phase from a sinusoidal
to a linear form. The latter represents the typical response of
a pair of classicaly anti-correlated spin-$\mathbf{J}$ systems,
analyzed through a dichotomic ``which emisphere'' measurement
\cite{Redh89, Evdo96}, i.e. the measurement of the angular momentum sign.
Such detection scheme is completely analogous to the dichotomic strategy
analyzed in this section.

The transition with increasing $n$ towards a classical response
for the singlet spin-$\frac{n}{2}$ can be explained observing
that the chosen dichotomic detection scheme is no more sufficient
to fully characterize the singlet spin states of increasing size
$n>1$. This measurement lacks of the necessary resolution \cite{Cost09}
to observe the peculiar quantum properties of these states. In other
words, this measurement scheme is not sufficient to fully extract
the information encoded in the polarization anti-correlation of the
singlet spin states. Their characterization would require a more
sophisticated detection apparatus able to discriminate the value
$m$ of the spin projection, i.e. in our case the difference in the
orthogonally polarized photon number, and not only its sign. An example
of such measurement \cite{Pere93} is given by the parity operator
$\hat{P}_{\pi, \pi_{\bot}} = \sum_{m=0}^{n} (-1)^{m} \vert (n-m)\pi,
m \pi_{\bot} \rangle \, \langle (n-m)\pi, m\pi_{\bot} \vert$.

The correlation between the two spatial modes of the singlet spin-$\frac{n}{2}$
states evaluated with this measurement operator leads to the following expression:
\begin{equation}
\begin{aligned}
P_{\vert \psi_{n}^{-} \rangle}(\theta) &= \langle \psi_{n}^{-} \vert \left(
\hat{P}_{+,-} \right)_{A} \otimes \left( \hat{P}_{\theta, \theta_{\bot}} \right)_{B}
\vert \psi_{n}^{-} \rangle \\ &= (-1)^{n} \frac{\sin\left[ (n+1) \theta \right]}
{(n+1) \sin \theta}
\end{aligned}
\end{equation}
This correlation function violates a CHSH inequality of an amount $S_{CHSH} = 2.481 > 2$
\cite{Pere93} even in the asymptotic limit of large number of particle
($n \rightarrow \infty$). However, such scheme based on the parity operator
requires a sharp photon number measurement in order to discriminate with unitary
efficiency among contiguous values of the spin projection.

\begin{widetext}

\begin{figure}[ht]
\centering
\includegraphics[width=0.95\textwidth]{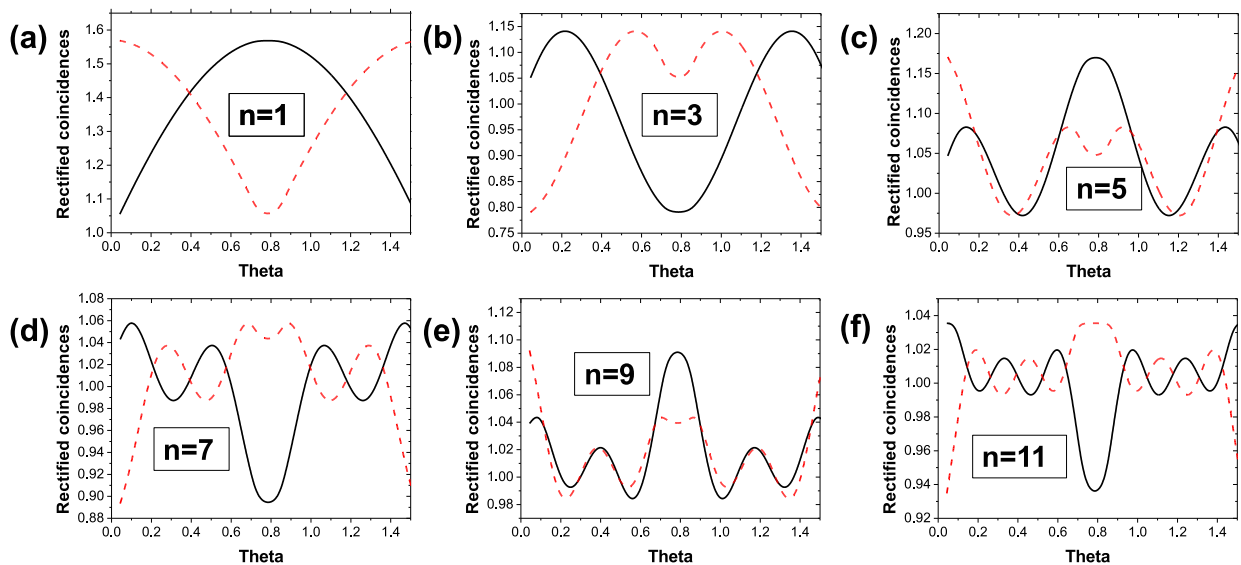}
\caption{(Color online) Plot of the interference fringe pattern $D^{(\pm 1, \pm
1)}_{\vert \protect\psi^{-}_{n} \rangle}(\protect\theta)$ for
singlet spin- $\frac{n}{2} $ states divided by a linear function
$L(\protect\theta)$ corresponding to the behaviour of two distinct
classical macroscopic objects.  Black continuous lines correspons to
the coincidences of both the (+1,+1) and (-1,-1)  outcome configurations,
while red dashed lines correspond to the (+1,-1) and (-1,+1) outcomes on the
two spatial mode. }
\label{fig:fourier_spont}
\end{figure}

\end{widetext}

As a further analysis, let us plot (Fig.\ref{fig:fourier_spont}) the
function $D^{(\pm1,\pm1)}_{\vert \psi^{n}_{-} \rangle}(\theta)/L(\theta)$,
which corresponds to the ratio between the interference fringe pattern of the
macro-macro configuration and a linear function of $\theta$. The choice of the
curve $L(\theta)$ as a reference is motivated by the following consideration.
The evaluation of the CHSH parameter in a system characterized by the linear response
leads to the maximum value in a classical framework $S_{CHSH} = 2$. Hence, this function $L(\theta)$
can be considered as the boundary between the ``classical'' and the ``quantum'' regions, since
it represents the response of two classical anti-correlated systems to this test.
In Fig.\ref{fig:fourier_spont}, we note that the ratio $D^{(\pm1,\pm1)}_{\vert \psi^{n}_{-} \rangle}(\theta)/L(\theta)$ presents
a number of intersections with the axis $y=1$ (unitary ratio) proportional to the value of $n$.
This depends on the explicit functional form of the interference fringe pattern of
Eq.(\ref{eq:int_fringe_pattern_n}).
Indeed, analyzing the explicit expression (Eq.(\ref{eq:coeffic_rotated})) of the coefficients $\epsilon_{m,p}^{n}(\theta)$, we find a sum
of terms $\left( \cos \theta \right)^{m+p-2q} \, \left( \sin \theta \right)^{n-m-p+2q}$, where the sum of the exponents is equal to the number
of photons $n$. Hence, the fringe pattern $D^{(\pm1,\pm1)}_{\vert \psi^{n}_{-} \rangle}(\theta)$
(Eq.(\ref{eq:int_fringe_pattern_n})) can be
re-organized in a Fourier series expansion containing all the Harmonics up to $k = 2n$. With increasing $n$,
the difference between $D^{(\pm1,\pm1)}_{\vert \psi^{n}_{-} \rangle}(\theta)$ and the linear function $L(\theta)$
is progressively reduced, since more harmonics are present in the Fourier expansion which asymptotically
reaches the expansion of $L(\theta)$.

In conclusion, the increase in the number of photons renders the dichotomic
measurement inefficient for the complete characterization of the state, and the
decreased correlations become more similar to classical ones.

\subsection{Propagation over a lossy channel}

Decoherence in macroscopic systems represents the main cause for
the impossibility of observing quantum phenomena in every-day life
and for the realization of quantum experimental schemes. In the
previous section we described a correlation experiment based on
dichotomic detection in an ideal setup, where both the
transmission channel and the detection apparatus possess unitary
quantum efficiency. However, in real setups loss in both stages
must be considered. Hence, the investigation on the effects of
decoherence allows to understand both the transition from the
quantum to the classical world and the feasibility of the schemes
here presented in order to violate the Bell's inequalities.

\begin{figure}[ht]
\centering
\includegraphics[width=0.4\textwidth]{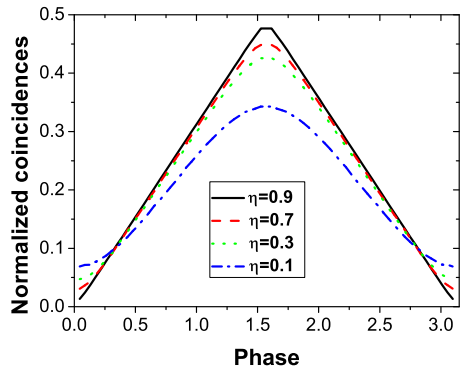}
\caption{(Color online) Interference fringe pattern in presence of losses for the singlet
spin-$\frac{n}{2}$ state for $n=51$. The effect of the lossy channel is the
progressive lowering of the visibility, while the dependence on the phase
changes from a linear to a sinusoidal pattern.}
\label{fig:fringe_losses_fixed_n=51}
\end{figure}

To this end, we introduce a beam-splitter model \cite{Loud00,Leon93} to
simulate losses phenomena on the macroscopic field here analyzed. In
particular, the analysis has been performed for any singlet spin-$\frac{n}{2%
}$ state which, in the decoherence-free case, exhibits the
transition from a sinusoidal interference pattern ($n=1$) to an
asymptotic linear interference fringe pattern ($n\gg 1$). Again, the
detection scheme used is a pure dichotomic measurement apparatus
(corresponding to the threshold detector and a O-Filter with
$h=k=0$). As the measurement operators of Eqs.
(\ref{eq:O-Filtering_POVM_1}-\ref{eq:threshold_POVM_3}) are linear
combination of Fock-state projectors, the lossy channel can be
simulated numerically.
More specifically, our calculation is divided in the following steps.
First, for a set of angles $\theta$ of the analysis basis on spatial mode
$\mathbf{k}_{B}$, we calculated the coefficients of the state $\epsilon^{n}_{m,p}(\theta)$
defined in Eq.(\ref{eq:coeffic_rotated}).
As the measurement operators are diagonal in the Fock basis, the results of
the measurement depend only on the diagonal part of the density matrix after losses.
Furthermore, the map that describes the lossy process, i.e. $\mathcal{L}[\hat{\rho}] =
\sum_{k} \gamma_{k} \hat{a}^{k} \hat{\rho} \hat{a}^{\dag \, k} \gamma^{\dag}_{k}$
where $\gamma_{k}= \frac{1}{\sqrt{k!}} (1-\eta)^{k/2} \eta^{(\hat{a}^{\dag} \hat{a})/2}$,
maps diagonal elements in diagonal elements of the density matrix. All these considerations
allow us to focus our numerical analysis on the diagonal part of the density matrix.
The numerical simulation proceeds as follows. Let us for example focus our attention on
the (+1,+1) joint outcome. The coefficients of the distribution $\vert \epsilon_{m,p}^{n}
(\theta) \vert^{2}$ are arranged in a matrix form, labelled by the row and column indexes $(m,p)$.
For each value of $m$ and $p$, 4 binomial random number generators with average
values respectively $\left\{ (n-m)\eta, m \eta, p \eta, (n-p)\eta \right\}$ simulate
a single shot passage of the $\vert (n-m)+, m -\rangle_{A} \otimes \vert p\theta,
(n-p) \theta_{\bot}\rangle_{B}$ element through the lossy channel of efficiency $\eta$.
Then, the output number of photons transmitted by the channel on each spatial mode
are dichotomically compared, assigning $1$ to the event if it belongs to the (+1,+1)
joint outcome configuration and $0$ otherwise. We then repeat this procedure $N$
times, averaging the results of the simulation for each value $m$ and $p$,
thus generating a matrix $M^{(+1,+1)}_{m,p}$ containing both the transmission and the
measurement processes. Finally, the point $D_{\hat{\rho}_{n, \eta}^{-}}^{(+1,+1)}(\theta)$
is reconstructed combining the matrix $M^{(+1,+1)}_{m,p}$ that describes the dinamical process
and the original photon number distribution of the state in the chosen basis, according to:
\begin{equation}
D_{\hat{\rho}_{n, \eta}^{-}}^{(+1,+1)}(\theta) = \sum_{m,p=0}^{n} M^{(+1,+1)}_{m,p} \vert
\epsilon_{m,p}^{n}(\theta)\vert^{2}
\end{equation}
The same procedure is applied for the other three outcomes of the joint dichotomic
measurements.
The results of the simulation for different values of the channel
transmittivity $\eta$ are reported for a fixed value of $n=51$ in
Fig.\ref{fig:fringe_losses_fixed_n=51}. As the efficiencies of the
channel and the detection scheme decrease, the linear fringe
patterns progressively evolve into sinusoidal ones, at the cost of
a smaller value of the visibility.

\subsection{O-Filtering and Threshold detection in a lossy regime}
\label{sec:OF_TD_fixed_n}

The analysis performed on the correlations present in the singlet spin-$%
\frac{n}{2}$ states has been focused on the detection by a pure dichotomic
scheme. We are now interested in observing the effects of more sophisticated
POVM measurements, as the threshold detection or the O-filtering methods
introduced in Section \ref{sec:Dichotomic_meas}. In particular, we analyze how
both the visibility and the form of the fringe pattern are modified exploiting
this different measurement schemes. The main idea beyond this approach concerns
the possibility of beating the losses effects on the macro-macro correlations,
by using a more sophisticated measurement than a pure dichotomic one.

We first analyze the correlations obtained by the O-Filtering detection
scheme, introduced in Section \ref{subsec:o-filtering_detection}. The fringe
pattern can be calculated by evaluating the average:

\begin{equation}
F^{(\pm 1, \pm 1)}_{\vert \psi^{-}_{n} \rangle}(\theta,h) = \left \langle
\left( \hat{F}^{(\pm 1)}_{+,-}(h) \right)_{A} \otimes \left( \hat{F}^{(\pm
1)}_{\theta, \theta_{\bot}}(h) \right)_{B} \right \rangle
\end{equation}

\begin{figure}[ht]
\centering
\includegraphics[width=0.5\textwidth]{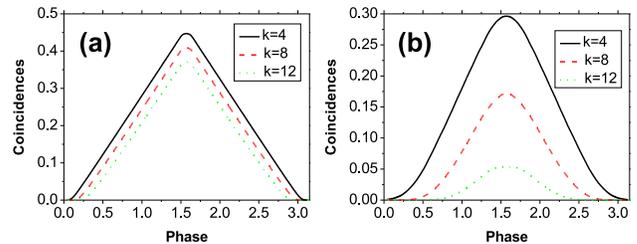}
\caption{(Color online) Effect of the O-Filtering detection technique on the fringe
pattern of a $n=51$ singlet state. (a) Transmittivity $\eta=1$ and
(b) transmittivity $\eta=0.3$. As the threshold $k$ is increased,
the tails of the fringe pattern are rounded.}
\label{fig:OF_fringe}
\end{figure}

\begin{figure}[ht]
\centering
\includegraphics[width=0.4\textwidth]{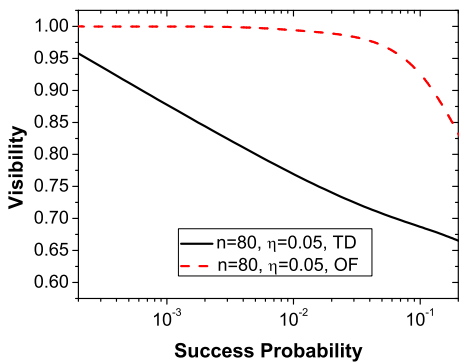}
\caption{(Color online) Trend of the visibility for the singlet spin states for $n=80$
and $\eta = 0.05$. The black straight curve corresponds to the TD detection
scheme, while the red dashed line to the OF apparatus. In both cases,
the success probability is calculated as the sum of the rate for the
two conclusive outcomes (+1) and (-1).}
\label{fig:OF_TD_visibility_n=80}
\end{figure}

We performed a numerical simulation, in order to consider also the
transmission over a lossy channel, with a procedure analogous to the one
described in the previuos section. We report in Fig.\ref{fig:OF_fringe}
the fringe pattern obtained for the $n=51$ singlet states for two values
of the channel efficiency. We note that, as the OF threshold $k$ is increased,
the tails of the fringe pattern are damped, while the form of the fringe
around the peaks remains unchanged. Furthermore, both the minimum and the
maximum of the fringes are lowered by this filtering procedure. To understand
the advantage of this measurement scheme with respect to the pure dichotomic case,
we analyze in Fig.\ref{fig:OF_TD_visibility_n=80} the trend of visibility of the fringe
pattern as a function of the threshold. We note that, for increasing $k$, the visibility
is increased by the filtering process. This advantage obtained by exploiting the O-filtering
measurement can be explained by the following considerations. In absence of
losses, the visibility of the fringe pattern is always unitary, as the analyzed state presents perfect
polarization anti-correlations. After the transmission over a lossy channel, the binomial
statistics added to the photon number distribution is responsible for the partial cancellation
of this property.
More precisely, if the difference between $n_{\pi}$ and $m_{\pi_{\bot}}$ on any of the two
spatial mode is little, losses may invert the outcome of a dichotomic measurement, i.e. for
example the (+1) outcome may be converted to the (-1) outcome if unbalanced losses occur
in that specific event. Such a process can generate the occurrence in the joint measurement
of a result with positive correlations, i.e. (+1,+1) or (-1,-1),  where in the decoherence-free
case only anti-correlations are present. Thus, the visibility of the fringe pattern
can be reduced by the presence of losses. However, to invert the outcome
of matrix elements with $n_{\pi} - m_{\pi_{\bot}} = q \gg 0$, a strongly
unbalanced losses in a single shot for the two polarization modes must occur. This event
has a decreasing probability as the difference $q$ becomes larger. Since the O-filter device
selects these zones of the Fock space which present such unbalancement, the outcome inversion
becomes practically neglible and the visibility of the fringe pattern progressively
returns unitary as the threshold $k$ is increased.

\begin{figure}[ht]
\centering
\includegraphics[width=0.5\textwidth]{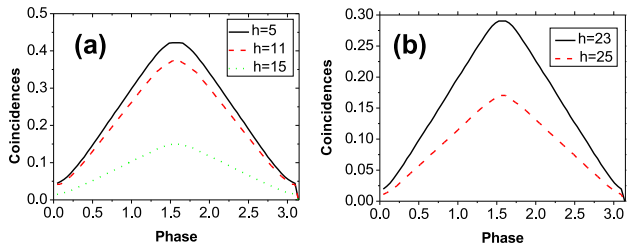}
\caption{(Color online) Effect of the Threshold detection technique on the fringe pattern
for a $n=51$ singlet state. (a) Transmittivity $\eta=0.3$ and
(b) transmittivity $\eta=0.5$. As the threshold $h$ on the total photon-number
is increased, the fringe patterns progressively return to have approximately
a linear dependence from the phase $\protect\theta$, as for the original $%
n=51$ singlet state. The values of the thresholds are indicated in the
figure.}
\label{fig:TH_fringe}
\end{figure}

Let us now consider the second POVM dichotomic measurement under investigation,
the threshold detection TD. The interference fringe pattern with this measurement
scheme can be calculated as:
\begin{equation}
T^{(\pm 1, \pm 1)}_{\vert \psi^{-}_{n} \rangle}(\theta,k) = \left \langle
\left( \hat{T}^{(\pm 1)}_{+,-}(k) \right)_{A} \otimes \left( \hat{T}^{(\pm
1)}_{\theta, \theta_{\bot}}(k) \right)_{B} \right \rangle
\end{equation}
In this expression, as before, the average is evaluated over the density matrix of the
state after the numerical simulation of the lossy channel. In Fig.\ref{fig:TH_fringe}
we report the form of the fringe pattern for $n=51$ and two different values
of the transmittivity of the channel. As the threshold $h$ is increased, we note that the
TD device is responsible for the progressive return of the fringe patterns to their
original form in absence of losses, i.e. for high values of $n$ an approximately linear form.
This behaviour can be explained as follows. While the original singlet-state has
a well definite number of photons, the lossy channel reduces the number of photons
to an average of $\eta \langle n \rangle$, with Poissonian fluctuations. At the
measurement stage the threshold $h$ in the TD device neglects
(Fig.\ref{fig:measurement_schemes}-(b))
the sectors of the Fock-space corresponding to a low number of photons. As $h$
approaches the value $h=n$, only the events in which the original singlet state
travels undisturbed in the channel (with probability $\eta^{2n}$) are selected, thus
restoring the original correlation.
We then analyze the effects of this measurement scheme in the visibility of
the fringe pattern in Fig.\ref{fig:OF_TD_visibility_n=80}. We note that this quantity
increases with a slower rate with respect to the OF apparatus. Differently
from the O-filtering case, on each spatial mode the zones of the Fock space
in which $n_{\pi} - m_{\pi_{\bot}}$ is small are not filtered out, and the increase
in the visibility is then much slower with the threshold $h$. However, also with
the TD apparatus the visibility reaches asymptotically the unitary value, since as said for
a threshold $h=n$ only the original singlet state, having unitary visibility, is detected.

The analysis carried in this section than shows that both the OF and the TD
detection strategies can be used to enhance the fringe pattern visibility
in lossy conditions for the singlet spin-$\frac{n}{2}$ states. A comparison
between the two schemes shows the better enhancement achievable with the OF
device. In Sec.\ref{sec:SPDC_theo}, we shall discuss in details the feasibility
of a CHSH test with such measurements.

\subsection{Investigation of non-locality with a CHSH-type inequality}
\label{sec:SPDC_theo}

In the previous paragraphs of this Section we reported the
interference fringe pattern obtained evaluating the correlations
between singlet spin-$\frac{n}{2}$ states both in absence and in
presence of experimental imperfections. The scheme here presented
can be exploited to perform a CHSH test \cite{Clau69} to investigate non-local
effects in these multiparticle states.

Let us briefly summarize in the light of a local hidden variable (LHV)
theory the content of Bell inequalities for a set of dichotomic
observables, by generalizing further the results already obtained by Reid
\textit{et al.} \cite{Reid02}. Consider a quantum state described by
the density matrix $\hat{\rho}$ defined in the Hilbert space $\mathcal{H}%
_{1}\otimes \mathcal{H}_{2}$. Define $\hat{O}_{a}^{i}$ the positive operator
acting on subspace $\mathcal{H}_{1}$, and the probability of finding the
value $i$ after the measurement $a$ as given by $\mathrm{Tr} \left[\hat{\rho}
(\hat{O}_{a}^{i}\otimes \hat{I}) \right]$.
The same relation holds for the positive operator $\hat{O}_{b}^{j}$ acting on
subspace $\mathcal{H}_{2}$.

The existence of a LHV model implies that the expectation values of the $a$
and $b$ are predetermined by the value of the parameter $\lambda $: \{$%
X_{a},X_{a^{\prime }},X_{b},X_{b^{\prime }}\},$ hence the product $a\cdot b$
is equal to $X_{a}(\lambda )X_{n}(\lambda )$. For a fixed value of $\lambda $
the variables $X_{n}$ with $n=\{a,b,a^{\prime },b^{\prime }\}$ take the
values ${-1,1}$ and satisfy the CHSH inequality:

{\small
\begin{equation}
X_{a}(\lambda )X_{b}(\lambda )+X_{a}(\lambda )X_{b^{\prime }}(\lambda
)+X_{a^{\prime }}(\lambda )X_{b}(\lambda )-X_{a^{\prime }}(\lambda
)X_{b^{\prime }}(\lambda )\leq 2  \label{eq:random_inequality}
\end{equation}
} The same inequality holds by integrating this equation on the space of the
hidden variable $(\lambda )$:

\begin{eqnarray}
&\,&\int_{\Omega }d\mathbb{P}(\lambda )X_{a}(\lambda )X_{b}(\lambda
)+\int_{\Omega }d\mathbb{P}(\lambda )X_{a}(\lambda )X_{b^{\prime }}(\lambda
)+  \notag  \label{eq:random_inequality_integrated} \\
&\,&\int_{\Omega }d\mathbb{P}(\lambda )X_{a^{\prime }}(\lambda
)X_{b}(\lambda )-\int_{\Omega }d\mathbb{P}(\lambda )X_{a^{\prime }}(\lambda
)X_{b^{\prime }}(\lambda )\leq 2  \notag \\
&&
\end{eqnarray}%
where $P(\lambda )$ is the measure of the $\lambda $ probability space. If
there is a local hidden variables model for quantum measurement taking
values $[-1,+1]$, then the following inequality must be satisfied by the measured mean values:
\begin{equation}
S_{CHSH} = E^{\rho }(a,b)+E^{\rho }(a,b^{^{\prime }})+E^{\rho }(a^{^{\prime
}},b)-E^{\rho }(a^{^{\prime }},b^{^{\prime }})\leq 2
\label{eq:CHSH_inequality}
\end{equation}%
where  $E^{\rho }(a,b)$ can be expressed as a function of the LHV as $E^{\rho }(a,b)=\int_{\Omega }X_{a}(\lambda )X_{b}(\lambda )d\mathbb{P}%
(\lambda )$. The violation of (\ref{eq:CHSH_inequality}) proves that a LHV
variables model for the considered experiment is impossible.

\begin{figure}[ht]
\centering
\includegraphics[width=0.4\textwidth]{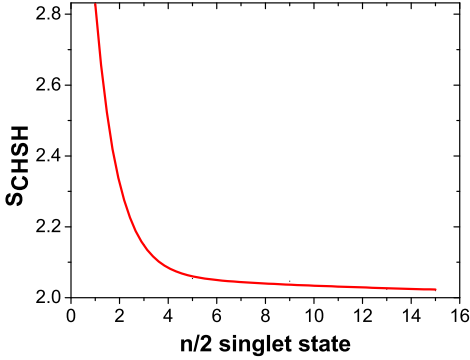}
\caption{(Color online) Value of the CHSH parameter $S^{\vert \psi^{-}_{n} \rangle}_{CHSH}$ for singlet spin-$\frac{n}{2}
$ states for an optimal choice of the angle settings and the dichotomic
``majority-voting'' measurement. We observe the progressive decrease in the
amount of violation for an increasing value of the number of photons present
in the state.}
\label{fig:CHSH_spont_optimized}
\end{figure}

In our case, the positive operators $\hat{O}_{a(b)}^{i}$ are given
by the dichotomic measurement operators $\left\{ \hat{T}^{(\pm 1)}_{\pi,
\pi_{\bot}}(0), \hat{F}^{(\pm 1)}_{\pi, \pi_{\bot}}(0) \right\}$.
In order to theoretically investigate the feasibility of a CHSH test
on the spin-$\frac{n}{2}$ states, we evaluated the $S_{CHSH}$
parameter in such system. The value of the $S_{CHSH}$ has been
numerically maximized over the measurement angles $\left\{ \theta,
\theta^{\prime},\varphi, \varphi^{\prime}\right\} $ of Alice's
[$\mathbf{a}(\theta)$ or $\mathbf{a}^{\prime }(\theta^{'})$] and Bob's
[$\mathbf{b}(\varphi)$ or $\mathbf{b}^{\prime }(\varphi^{'})$]
the polarization basis. In Fig.\ref {fig:CHSH_spont_optimized} we report
the results obtained for different values of the number of
photons, and hence the spin, of the analyzed state. We observe the
decrease in the absolute value of $S^{\vert \psi^{-}_{n} \rangle}_{CHSH}$ analogously to what
reported in \cite{Reid02,Banc08} for an equivalent Bell's
inequalities test. However, the asymptotic behavior for high $n$
shows that the parameter $S_{CHSH}$ never falls below the
classical limit, but the amount of violation progressively becomes
smaller and any decoherence process may forbid its experimental
observation.

We now discuss the feasibility of a Bell's inequality test
when the OF and the TD detection methods are adopted in the context
of Local Hidden Variables (LHV) models. This analysis is motivated
by the increase in the visibility obtained with this measurement
operators with respect to the pure dichotomic case. Both strategies
present the POVM feature of having three possible outcomes $\left\{
-1, 1, 0 \right\}$, at variance with a genuine dichotomic
measurement. In order to clarify the validity of a Bell test in presence
of such kind of POVM's, let us consider the case in which at the
A site a standard dichotomic measurement is performed, while at
the B site a POVM measurement is carried out.

Consider the outcomes for which the Bob's
results are different from $0$. In this case the expectation value of the
product of $a$ and $b$ is conditioned by the event:
``outcome $b$ different from zero''. In a LHV model these
conditional expectations are represented by:
\begin{equation}
E^{\rho }(a\cdot b)=\int_{\Omega ^{^{\prime }}}X_{a}(\lambda )X_{b}(\lambda
)d\mathbb{P^{\prime }}(\lambda )
\end{equation}%
where $\Omega ^{^{\prime }}$ is the hidden variable probability sub-space
for which, for \textit{any} $X_{b^{\prime }}(\lambda )$, $\ $is $%
X_{b}(\lambda )\neq 0$ and $d\mathbb{P^{\prime }}=d\mathbb{P}/\int {_{\Omega
^{^{\prime }}}d\mathbb{P}}$. Similarly:

\begin{equation}
E^{\rho }(a\cdot b^{\prime })=\int_{\Omega ^{^{\prime \prime
}}}X_{a}(\lambda )X_{b}(\lambda )d\mathbb{P^{\prime \prime }}(\lambda )
\end{equation}
where $\Omega ^{^{\prime \prime }}$ is the hidden variable probability
sub-space for which, for \textit{any} $X_{b}(\lambda )$, is $X_{b^{\prime
}}(\lambda )\neq 0$ and $d\mathbb{P^{\prime \prime }}=d\mathbb{P}/\int {%
_{\Omega ^{^{\prime \prime }}}d\mathbb{P}}$. Since for different random
variables $X_{b}$ and $X_{b^{\prime }}$ these conditional expectations
values can in principle refer to different subensembles $\Omega ^{\prime }$
and $\Omega "$ of the original ensemble $\Omega $, in general the equation (%
\ref{eq:random_inequality_integrated}) doesn't hold any more and the
measured quantity, based on the detection of conditional values, is:

\begin{eqnarray}
&&\int_{\Omega ^{\prime }}d\mathbb{P^{\prime }}(\lambda )X_{a}(\lambda
)X_{b}(\lambda )+\int_{\Omega ^{\prime }}d\mathbb{P^{\prime }}(\lambda
)X_{a^{\prime }}(\lambda )X_{b}(\lambda )+  \notag
\label{eq:random_inequality_without_assumption} \\
&&\int_{\Omega ^{\prime \prime }}d\mathbb{P^{\prime \prime }}(\lambda
)X_{a}(\lambda )X_{b^{\prime }}(\lambda )-\int_{\Omega ^{\prime \prime }}d%
\mathbb{P^{\prime \prime }}(\lambda )X_{a^{\prime }}(\lambda )X_{b^{\prime
}}(\lambda )  \notag \\
&&
\end{eqnarray}

Let us consider the class of LHV models such that, for a fixed value of $%
\lambda $, simultaneously is: $X_{b}(\lambda )\neq 0,X_{b^{\prime }}(\lambda
)\neq 0$. In this case the inequality (\ref{eq:random_inequality}) still
holds since it becomes:
\begin{eqnarray}
&&\int_{\Omega \ast }d\mathbb{P^{\ast }}(\lambda )X_{a}(\lambda
)X_{b}(\lambda )+\int_{\Omega \ast }d\mathbb{P^{\ast }}(\lambda
)X_{a}(\lambda )X_{b^{\prime }}(\lambda )+  \notag
\label{eq:random_inequality_POVM} \\
&&\int_{\Omega \ast }d\mathbb{P}^{\ast }(\lambda )X_{a^{\prime }}(\lambda
)X_{b}(\lambda )-\int_{\Omega \ast }d\mathbb{P^{\ast }}(\lambda
)X_{a^{\prime }}(\lambda )X_{b^{\prime }}(\lambda )\leq 2  \notag \\
&&
\end{eqnarray}%
where $\Omega ^{\ast }$ is the hidden variable probability common subspace
for which $X_{b}(\lambda )\neq 0$ and $X_{b^{\prime }}(\lambda )\neq 0$.%
\newline
With reference to our experimental situation, let us now make an auxiliary assumption implying that the probability of rejecting a
measurement\ does not depend on the hidden parameter $\lambda $ and on the
measurement settings, i.e. $\Omega ^{\prime }=\Omega ^{\prime \prime
}=\Omega ^{\ast }$ \cite{Aden03}. In this case the experimentally observed
quantity (\ref{eq:random_inequality_without_assumption}) will follow the LHV
inequality (\ref{eq:random_inequality_POVM}), and its violation implies the
non-locality of the considered system. While for what concerns the OF based strategy this assumption on the LHV is a strong one,
in the TD case it is legitimated by the fact that
the Hilbert subspace leading to a conclusive outcome is
invariant under any rotation of the polarization basis since data are excluded depending only on the overall number of photons. In other words, when an
event leads to a $(\pm 1)$ outcome for a specific choice of the measurement basis, it would correspond to a conclusive outcome if measured in another basis. This scenario is exactly the same encountered in any two-photon Bell inequality test and hence requires a fair sampling assumption.

To conclude the discussion, we briefly analyze the advantages of the two
POVM schemes presented here in terms of the achievable violation of
the CHSH inequality $S_{CHSH}-2$.
In the OF case, we expect that the fast increase in the visibility
may lead to an increase in the amount of violation with respect to the
pure dichotomic measurement. In the TD case, as already discussed in the previous
section, the effect of the threshold $h$ is the restoration of
the original correlations present in the $\vert \psi^{-}_{n} \rangle$
state before the lossy channel. This means that the value of
the $S_{CHSH}$ parameter reaches for $h=n$ the maximum value $S^{\vert \psi^{-}_{n}
\rangle}_{CHSH}$, reported in Fig.\ref{fig:CHSH_spont_optimized}, and
the amount of achievable violation becomes practically negligible for large $n$.

\subsection{Spontaneous Parametric Down Conversion: interference fringe pattern}

The following step of our theoretical analysis is the investigation on the interference
fringe pattern obtained by the process of spontaneous parametric down conversion
exploiting the dichotomic measurement schemes presented in Sec.\ref{sec:Dichotomic_meas}.
As already stressed in Eqs.(\ref{eq:SPDC_state}-\ref{eq:singlet_n}), this optical source
generates a quantum superposition of the singlet spin-$\frac{n}{2}$ states.
We performed the same calculation of Sec.\ref{sec:OF_TD_fixed_n} in order to analyze both the form
of the interference fringe pattern and the trend of the visibility when the two
dichotomic measurements (OF and TD) were exploited at the detection stage.

\begin{figure}[ht]
\centering
\includegraphics[width=0.5\textwidth]{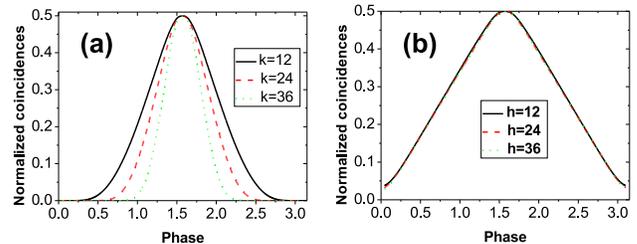}
\caption{(Color online) Fringe pattern in presence of losses for the SPDC output state
with the two different analyzed dichotomic measurements. In all curves, $g=2.5$ and
$\eta = 0.5$. (a) Normalized
fringe pattern with the O-filter device for different values of the threshold $k$.
The three curves correspond to a filtering signal of $P(k=12) = 0.116$ (black straight curve),
$P(k=24) = 0.014$ (red dashed curve) and $P(k=36) = 2.07 \times 10^{-4}$ (green dotted curve). We note
the increase in the smoothing of the minimum of the fringes. (b) Normalized
fringe pattern with the TD device for different values of the threshold $h$.
The three curves correspond to a filtered signal of $P(h=12) = 0.328$ (black straight curve),
$P(h=24) = 0.068$ (red dashed curve) and $P(h=36) = 1.07 \times 10^{-3}$ (green dotted curve). We note
that the form of the normalized fringes is left unchanged by the TD device.}
\label{fig:fringe_SPDC_losses_OF_TH}
\end{figure}

We then report in Fig.\ref{fig:fringe_SPDC_losses_OF_TH} the form of the fringe pattern
for the SPDC output state in presence of losses, with $g=2.5$ and $\eta = 0.5$. Analogously to
what observed for singlet spin states, the effect of the two measurement devices is different.
On one side, the O-filtering technique is responsible for a smoothing of the fringe pattern
tails, while on the other side, the Threshold Detector leaves the form of the
fringe pattern unaltered.

\begin{figure}[ht]
\centering
\includegraphics[width=0.5\textwidth]{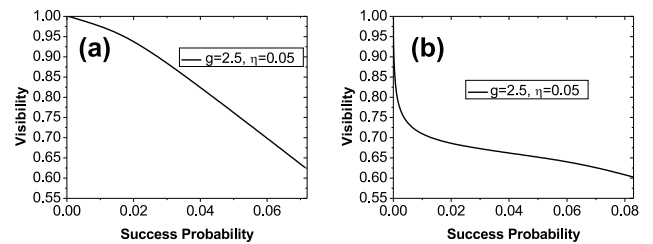}
\caption{(Color online) (a) Visibility of the fringe pattern as a function of the
success probability $\langle \hat{F}^{(+1)}_{\pi, \pi_{\bot}}(k)
\rangle + \langle \hat{F}^{(-1)}_{\pi, \pi_{\bot}}(k) \rangle$ for
the SPDC states analyzed with the OF device. (b) Visibility of the
fringe pattern as a function of the success probability $\langle
\hat{T}^{(+1)}_{\pi, \pi_{\bot}}(h) \rangle + \langle \hat{T}^{(-1)}_{\pi,
\pi_{\bot}}(h)\rangle$ for the SPDC states analyzed with the TD device.
For both curves, $g=2.5$ and $\eta = 0.05$.}
\label{fig:figure_SPDC_fit_OF_TD}
\end{figure}

To conclude the analysis of this section, we report in Fig.\ref{fig:figure_SPDC_fit_OF_TD} the trend
of the visibility of the fringe pattern as a function of the success probability of the
two dichotomic measurement devices. A comparison between the technique shows the faster
increase of the visibility in the OF case (Fig.\ref{fig:figure_SPDC_fit_OF_TD}-(a))
with respect to the TD one (Fig.\ref{fig:figure_SPDC_fit_OF_TD}-(b))
The cost of this faster increase in the visibility is a loss in the universality
of the device, since the Threshold Detector selects a regions of the Fock space which is
invariant under rotations of the polarization basis, at variance with the O-filter.

\section{Experimental observation of correlations in high gain SPDC}
\label{sec:experimental}

In order to complete our analysis on the
correlations connecting a macro-macro state obtained via high gain
optical parametric amplification, we have experimentally
investigated the conceptual scheme presented in the previous
sections. We have generated a multiphoton state through an EPR
source and we have performed dichotomic measurement via O-Filter (OF)
and Threshold Detector (TD) upon it. In this section we report
the experimental interference fringe patterns observed for the
spontaneous field generated by the high gain OPA working in a non
collinear configuration. As a first step we shall characterize the
OPA in a high gain regime, by evaluating the non linear gain of
the amplifier and by reporting the generated field fringe pattern
visibility as a function of the gain. Then, we shall investigate
the features of the multiphoton field through the two measurement
strategies studied in Sec. \ref{sec:Dichotomic_meas}.

\begin{widetext}

\begin{figure}[th]
\centering
\includegraphics[width=0.8\textwidth]{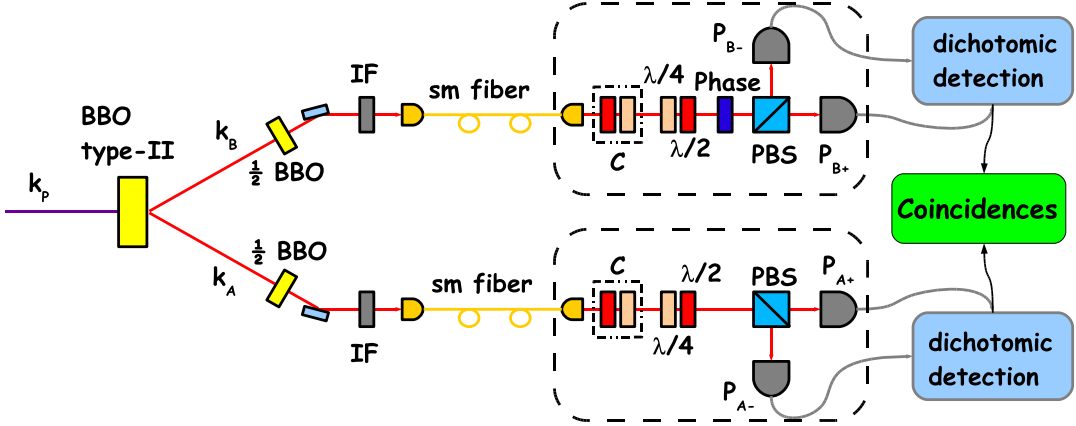}
\caption{(Color online) Experimental setup for the generation and detection of a bipartite
macroscopic field. The high laser pulse on mode $\mathbf{k}_{P}$ excites a
type-II EPR source in the high gain regime, i.e. $g=3.5$. The two spatial
mode $\mathbf{k}_{A}$ and $\mathbf{k}_{B}$ are spectrally and spatially
selected by interference filters (IF) and single mode fibers. After fiber
compensation $C$, the two modes are analyzed in polarization and detected by
four photomultipliers ($P_{A+}$,$P_{A-}$,$P_{B+}$,$P_{B-}$). The signals are
then analyzed electronically to perform either the threshold dichotomic
detection described in the paper or the Orthogonality filtering detection
technique. Finally, the coincidences between the measurement outcomes are
recorded to obtain the desired interference fringe patterns.}
\label{fig:experimental_setup_spont}
\end{figure}

\end{widetext}

Let us now describe the experimental setup shown in Fig.\ref
{fig:experimental_setup_spont}.
The excitation source was a Ti:Sapphire Coherent Mira mode-locked laser
amplified by a Ti:Sapphire regenerative RegA device operating with
repetition rate 250 kHz. The output beam, frequency-doubled by
second-harmonic generation, provided the OPA excitation field beam at the UV
wave-length (wl) $\lambda =397.5$ nm with power 600 mW on mode $\mathbf{k}_{P}
$.The SPDC source was a BBO crystal cut for type-II phase-matching, working
in a non-collinear configuration \cite{Kwia95}, in a high gain regime. The
evaluated non linear gain is $g=3.49\pm 0.05$ corresponding to the
generation of an average number of photons per mode of $\overline{n}\approx
270$ per pulse, corresponing to an overall average value of $\langle n\rangle \approx 540$
on each spatial mode.

The multiphoton fields on modes $\mathbf{k_{A}}$ and
$\mathbf{k_{B}}$ were filtered by $1.5$ nm interferential filters
(IF) and coupled by single mode fibers. The signals were then
attenuated, analyzed in polarization and detected by single photon
 SPCM detectors (not shown in Fig.\ref
{fig:experimental_setup_spont}).

\begin{figure}[ht]
\centering
\includegraphics[width=0.4\textwidth]{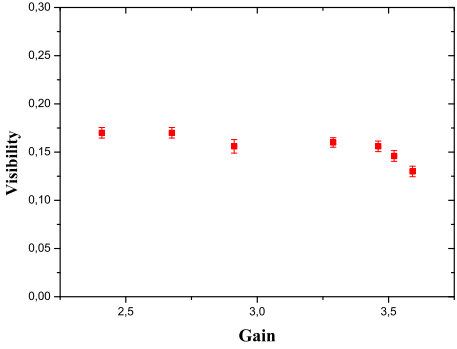}
\caption{(Color online) Experimental trend of the visibility as a function of the NL-gain. }
\label{fig:visi_vs_gain}
\end{figure}

In order to characterize the source, we performed a set of preliminary measurements
exploiting a SPCM detector on both spatial modes, deliberately attenuating the generated in
order to have only few photons incident on the detector. First, we
measured the non-linear gain of the amplifier studying how the detected
signal is increased by varying the power of the incident pump beam on
the crystal.
\begin{figure}[ht]
\centering
\includegraphics[width=0.4\textwidth]{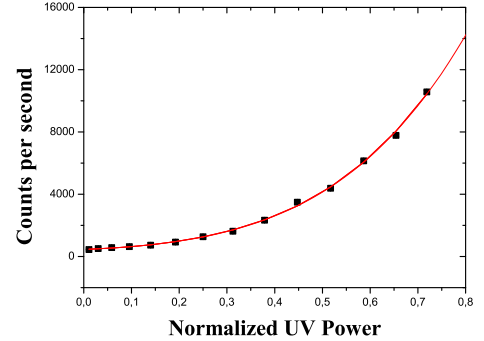}
\caption{(Color online) Experimental evaluation of the amplifier NL-gain: we report the
counts of an SPCM detector on mode $\mathbf{k_{A}}$ versus the normalized UV
power, defined as $I_{\mathrm{in}}/I_{\mathrm{max}}$. The red curve reproduces
the best fit of the experimental data, the
expected trend function is reported in \protect\cite{Eise04}.}
\label{fig:gain}
\end{figure}
In Fig.\ref{fig:gain} we report the counts registered on mode $\mathbf{k_{A}}
$ by a SPCM detector as a function of the normalized UV power signal. The
evaluation of the NL-gain has been performed as shown in Ref.\cite{Eise04},
and, as said, we found $g=3.49\pm 0.05$. As a further investigation on the
multiphoton field features, we registered the coincidences between the
signals on mode $\mathbf{k_{A}}$ and $\mathbf{k_{B}}$, as a function of the
phase $\varphi $, that represents the variation of the polarization analysis
basis on Bob site, i.e. $\vec{\pi} _{\varphi }=\vec{\pi}_{H}+e^{i\varphi} \vec{\pi}_{V}$.
Both fields are detected by two SPCM at Alice's and Bob's sites. Again,
the signals were attenuated in order to have few photons incident on the
detectors, in order to work in a linear response regime for the SPCM. The
visibility of the obtained fringe patterns as a function the NL-gain is
shown in Fig. \ref{fig:visi_vs_gain}. As stressed in \cite{Eise04}, the
trend of visibility decreases as the gain increases, this is due to
losses and to limited detectors photon number resolution. The decrease of
visibility below the theoretical asymptotic value of $33\%$ is due to the
multimodal operation of the amplifier, although, differently from what is
reported in \cite{Eise04}, we observe a value of visibility that remains
above $15\%$ as far as the NL-gain reaches the value of $3.5$, while in \cite
{Eise04} the visibility seems to fall below $15\%$ for gain values higher
than $2$.

\subsection{Non-collinear SPDC analyzed with the Orthogonality Filter}

In this Section we report the observation of the fringe patterns obtained by
the O-Filtering measurement strategy illustrated in Sec.\ref
{sec:Dichotomic_meas}-A.\newline
The multiphoton fields at Alice's and Bob's site are analyzed in
polarization and detected by two photomultipliers (PMs), $(P_{A+},P_{A-})$
and $(P_{B+},P_{B-})$ respectively. This devices produce on each pulse a macroscopic
output electronic current, whose amplitude is linearly proportional to the number
of incident photons.

\begin{figure}[ht]
\centering
\includegraphics[width=0.4\textwidth]{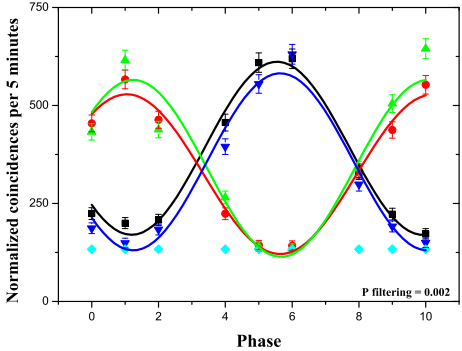}
\caption{(Color online) Fringe patterns obtained by filtering on the difference
of the signals. The main visibility is $0.67 \pm 0.02$. Coincidences
have been normalized to the product of the signals detected on
each of the analyzed outcomes of the OF.}
\label{fig:frange_filtraggio_diff}
\end{figure}

Let us fix the polarization analysis basis at Bob's site: the PMs provide
the electronic signals $(I_{+}^{B},I_{-}^{B})$ corresponding to the field
intensity on the mode $\mathbf{k}_{B}$ associated with the $\pi -$components
$(\overrightarrow{\pi }_{+},\overrightarrow{\pi }_{-})$, respectively. By
the OF, shot by shot the difference signals $\pm (I_{+}^{B}-I_{-}^{B})$ are compared with
a threshold $\xi k>0$, where $\xi$ is a constant describing the response
of the photomultipliers. When the condition $(I_{+}^{B}-I_{-}^{B})>\xi k$ \
is satisfied, a standard transistor-transistor-logic (TTL) electronic
square-pulse $L_{B}$ is realized at one of the two output ports of OF.
Likewise, when the condition $(I_{-}^{B}-I_{+}^{B})>\xi k$ is satisfied, a $%
L_{B}^{\ast }$ TTL pulse is realized at other output port of OF. The PM\
output signals are discarded for $-\xi k<(I_{+}^{B}-I_{-}^{B})<\xi k$, i.e.
in condition of low state discrimination. By increasing the value of $\ $%
the\ threshold $k$ an increasingly better discrimination is obtained
together with a decrease of the rate of successful detection. The same
measurement strategy is adopted at Alice's site, where the output TTL
signals $(L_{A},L_{A}^{\ast })$ are generated. The fringe patterns are
obtained by the following procedure: the analysis basis at Alice's site is kept
fixed while the basis at Bob's site is varied through an adjustable phase delay
given by a Babinet-Soleil compensator. Finally the coincidences
between the TTL signals at Alice's and Bob's site are taken into account,
namely $(L_{A},L_{B}),(L_{A},L_{B}^{\ast }),(L_{A}^{\ast
},L_{B}),(L_{A}^{\ast },L_{B}^{\ast })$. We report in figure \ref
{fig:frange_filtraggio_diff} the corresponding fringe patterns obtained in
the $\{\pi _{+},\pi _{-}\}$ basis, analogous results are observed in the $%
\{\pi _{R},\pi _{L}\}$ and $\{\pi _{H},\pi _{V}\}$ basis, due to
the irrotational invariance of the generated multiphoton state.
The threshold k was set so that the percentage of data taken into
account was $2\times 10^{-3}$ of the overall sample.

\begin{figure}[ht]
\centering
\includegraphics[width=0.4\textwidth]{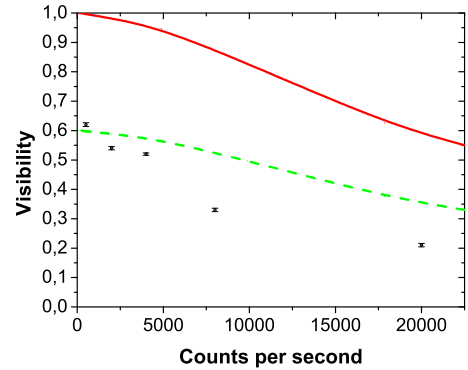}
\caption{(Color online) Trend of visibility versus OF counts. The theoretical predictions
(continuous red line) has been renormalized (dotted green line) respect to
the maximum reached visibility value.}
\label{fig:visi_vs_counts_OF}
\end{figure}

For sake of completeness we report the trend of visibility as a function of
the OF counts in Fig.\ref{fig:visi_vs_counts_OF}. We observe an increase of
visibility as the counts detected decrease. The highest
visibility obtained is not enough to violate the CHSH inequality, due to the
inefficiency of a dichotomic measurement performed on a multiphoton quantum
state and to experimental imperfections. However, in accordance with theoretical predictions, we observe that
the OF technique allows to minimize losses effects.

\begin{figure}[hb]
\centering
\includegraphics[width=0.4\textwidth]{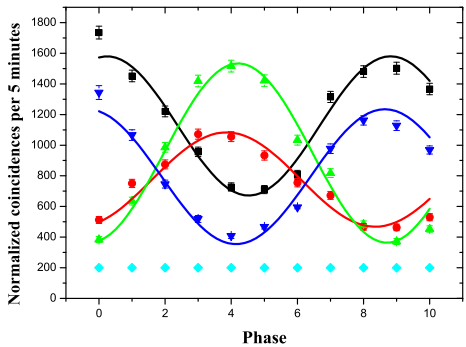}
\caption{(Color online) Fringe patterns obtained by filtering on the sum of the
signals. The main visibility is $0.49 \pm 0.02$.}
\label{fig:frange_filtraggio_somma}
\end{figure}

\subsection{Non-collinear SPDC analyzed with threshold detection}

\begin{figure}[ht]
\centering
\includegraphics[width=0.4\textwidth]{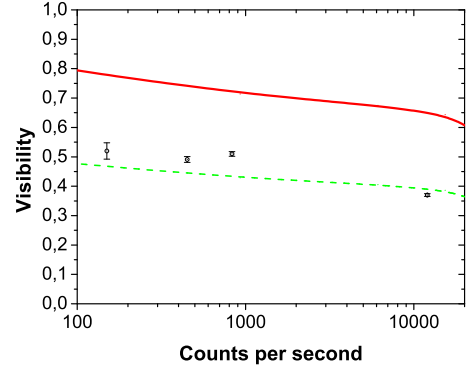}
\caption{(Color online) Visibility versus threshold detector counts.The theoretical predictions
(continuous red line) has been renormalized (dotted green line) respect to
the maximum reached visibility value.}
\label{fig:stampa_visi}
\end{figure}

A further investigation on the macro-macro correlation has been carried out
by performing another dichotomic measurement on the amplified states on
modes $\mathbf{k_{A}}$ and $\mathbf{k_{B}}$. The signals detected by the
photomultipliers $(P_{A+},P_{A-})$ and $(P_{B+},P_{B-})$ enter into two
threshold detectors (TD), that performs the shot by shot measurement illustrated in Sec.%
\ref{sec:Dichotomic_meas}-B. Each TD works as follows: the PMs electronic
signals $(I^{B}_{+},I^{B}_{-})$ ($(I^{A}_{+},I^{A}_{-})$) corresponding to
the field intensity on the mode $\mathbf{k}_{B}$ ($\mathbf{k}_{A}$),
associated with the $\pi -$components $(\overrightarrow{\pi }_{+},%
\overrightarrow{\pi }_{-})$ respectively, enter into the TD. By it the sum
signals $\pm (I^{B}_{+}+I^{B}_{-})$ ($\pm (I^{A}_{+}+I^{A}_{-})$) are
compared with a threshold $\xi h>0$ . When the condition $%
(I^{B}_{+}+I^{B}_{-})>\xi h$ and $I^{B}_{+}-I^{B}_{-}>0$ \ ($%
(I^{A}_{+}+I^{A}_{-})>\xi h$ and $I^{A}_{+}-I^{A}_{-}>0$) is satisfied, a
standard transistor-transistor-logic (TTL) electronic square-pulse $J_{B}$ ($%
J_{A}$) is realized at one of the two output ports of TD. On the other hand
when the condition $(I^{B}_{+}+I^{B}_{-})>\xi h$ and $I^{B}_{-}-I^{B}_{+}>0$
\ ($(I^{A}_{+}+I^{A}_{-})>\xi h$ and $I^{A}_{-}-I^{A}_{+}>0$) is satisfied,
a standard transistor-transistor-logic (TTL) electronic square-pulse $%
J_{B}^{*}$ ($J_{A}^{*}$) is realized at the other output ports of TD.
Finally the coincidences between signals $%
(J_{A},J_{B}),(J_{A},J_{B}^{*}),(J_{A}^{*},J_{B}),(J_{A}^{*},J_{B}^{*})$
are registered by a coincidences box. The obtained fringe patterns
corresponding to a detection probability equal to $P=1.6 \times
10^{-3}$ are shown in Fig.\ref {fig:frange_filtraggio_somma}.
Finally, a study on the obtained visibility as a function of the
fraction of considered data has been carried out. We report in
Fig.\ref{fig:stampa_visi} the trend of visibility versus TDs
counts.

\section{Conclusion and perspectives}

In this work we have reported a deep analysis on the possibility
of observing quantum correlation on a multiphoton quantum system
by performing probabilistic dichotomic measurements. We have
addressed a specific class of multiphoton states: the ones
obtained by the high gain optical parametric amplifier working in
a non-collinear configuration. To this end we have introduced two
kind of dichotomization processes, based on the O-Filtering
procedure discussed in \cite{DeMa08} and on a threshold detection
scheme similar to the naked eye discussed in \cite{Brun08}. It has
been demonstrated that these two detection schemes reduce to a
simple dichotomic measurement when their characteristic thresholds
are set to 0. We have shown that such dichotomic measurement when
performed on $\frac{n}{2}$-spin states with increasing $n$, asymptotically
allows in the ideal case the violation of CHSH Bell's inequality even for
large $n$. The shape of correlation functions has been investigated, and
we have shown that the sinusoidal correlation pattern, typical of an
$\frac{1}{2}$-spin state, tends asymptotically to a triangular
form, proper to classical correlations. When losses and
decoherence are introduced the visibility of the correlation
pattern is lowered and its shape turns out to be sinusoidal. In
presence of losses, the violation of CHSH Bell's inequality is not
allowed by a dichotomic measurement and more complicated detection
schemes are required. We then discussed in terms of LHV models
the feasibility of a CHSH test with the two probabilistic
measurements presented in this paper.

Finally, we have shown experimentally that the measurement
performed by the probabilistic dichotomic schemes, the O-Filter
and the Threshold Detector, allow to obtain higher visibility
of correlation functions, not enough to violate CHSH Bell's
inequality, but effective to reduce losses and decoherence
effects. An open question concerns the existence of entanglement criteria
able to demonstrate the presence of entanglement in a "macro-macro" scenario,
and the possibility of adopting the measurement devices introduced in this paper
within such context.

In conclusion we believe that our analysis contributes to shed light on the
role of measurement performed on large size quantum systems, and on the
possibility of observing entanglement and quantum phenomena at a macroscopic
level.

%

\begin{thebibliography}{10}

\bibitem{Bell64}
J.~Bell,
\newblock Physics {\bf 1}, 195 (1964).

\bibitem{Clau69}
J.~F. Clauser, M.~A. Horne, A.~Shimony, and R.~A. Holt,
\newblock Phys. Rev. Lett. {\bf 23}, 880 (1969).

\bibitem{Alle90}
Y.~H. Shih, and C.~O. Alley 
\newblock Phys. Rev. Lett. {\bf 61}, 2921 (1988).

\bibitem{Ou88}
Z.~Y. Ou and L.~Mandel
\newblock Phys. Rev. Lett. {\bf 61}, 50 (1988).

\bibitem{Kies98}
T.~E. Kiess, Y.~H. Shih, A.~V. Sergienko, and C.~O. Alley
\newblock Phys. Rev. Lett. {\bf 71}, 3893 (1993).

\bibitem{Chen06}
K. J. Resch, P. Walther, and A. Zeilinger
	\newblock Phys. Rev. Lett. {\bf 94}, 070402 (2005).

\bibitem{Walt05}
P.~Walther, M.~Aspelmeyer, K.~J. Resch, and A.~Zeilinger
\newblock Phys. Rev. Lett. {\bf 95}, 020403 (2005).

\bibitem{Zure03}
W.~H. Zurek,
\newblock Rev. Mod. Phys. {\bf 75}, 715 (2003).

\bibitem{Pere93}
A.~Peres,
\newblock {\em Quantum Theory: Concepts and Methods} (Kluwer, 1993).

\bibitem{Kofl07}
J.~Kofler and C.~Brukner,
\newblock Phys. Rev. Lett. {\bf 99}, 180403 (2007).

\bibitem{Kofl08}
J.~Kofler and C.~Brukner,
\newblock Phys. Rev. Lett. {\bf 101}, 090403 (2008).

\bibitem{Kofl09}
J.~Kofler, N.~Buric, and C.~Brukner,
\newblock arXiv.org:0906.4465  (2009).

\bibitem{Jeon09}
H.~Jeong, M.~Paternostro, and T.C.~Ralph,
\newblock Phys. Rev. Lett. {\bf 102}, 060403 (2009).

\bibitem{Chen02}
 Z.B.~Chen, J.W.~Pan, G. Hou, and Y.D. Zhang,
\newblock Phys. Rev. Lett. {\bf 88}, 040406 (2002).

\bibitem{Port06}
S.~Portolan, O.~{Di Stefano}, S.~Savasta, F.~Rossi, and R.~Girlanda,
\newblock Phys. Rev. A {\bf 73}, 020101(R) (2006).

\bibitem{Reid02}
M.~D. Reid, W.~J. Munro, and F.~{De Martini},
\newblock Phys. Rev. A {\bf 66}, 033801 (2002).

\bibitem{Banc08}
J.~D. Bancal,  C. Branciard,  N. Brunner, N. Gisin,
	 S. Popescu, and C. Simon
\newblock Phys. Rev. A {\bf 78}, 062110 (2008).

\bibitem{Simo03}
C. Simon, and D. Bouwmeester 
\newblock Phys. Rev. Lett. {\bf 91}, 053601 (2003).

\bibitem{Eise04}
H.~S. Eisenberg, G.~H. Khoury, G.~A. Durkin, C.~Simon, and D.~Bouwmeester,
\newblock Phys. Rev. Lett. {\bf 93}, 193901 (2004).

\bibitem{Cami06}
M.~Caminati, F.~{De Martini}, R.~Perris, F.~Sciarrino, and V.~Secondi,
\newblock Phys. Rev. A {\bf 73}, 032312 (2006).

\bibitem{Kwia95}
P.~G. Kwiat, K. Mattle, H. Weinfurter, A. Zeilinger,
	A.~V. Sergienko, and Y. Shih
\newblock Phys. Rev. Lett. {\bf 75}, 4337 (1995).

\bibitem{Gric97} 
W.P. Grice and I.A. Walmsley
\newblock Phys. Rev. A {\bf 56}, 1627 (1997)

\bibitem{Eibl03}
M.~Eibl, S. Gaertner, M. Bourennane, C.
	Kurtsiefer, M. Zukowski, and H. Weinfurter
\newblock Phys. Rev. Lett. {\bf 90}, 200403 (2003).

\bibitem{Wein01}
H.~Weinfurter and M.~Zukowski,
\newblock Phys. Rev. A {\bf 64}, 010102(R) (2001).

\bibitem{Wiec08}
W.~Wieczorek, C. Schmid, N. Kiesel, R.
	Pohlner, O. G\"{u}hne, and H. Weinfurter
\newblock Phys. Rev. Lett. {\bf 101}, 010503 (2008).

\bibitem{Naga07}
E.~Nagali, T.~{De Angelis}, F.~Sciarrino, and F.~{De Martini},
\newblock Phys. Rev. A {\bf 76}, 042126 (2007).

\bibitem{DeMa08}
F.~{De Martini}, F.~Sciarrino, and C.~Vitelli,
\newblock Phys. Rev. Lett. {\bf 100}, 253601 (2008).

\bibitem{Seka09}
P.~Sekatski, N.~Brunner, C.~Branciard, N.~Gisin, and C.~Simon,
\newblock Phys. Rev. Lett. {\bf 103}, 113601 (2009).

\bibitem{Redh89}
M.~Redhead,
\newblock {\em Incompleteness, Nonlocality, and Realism} (Oxford University
  Press, 1989).

\bibitem{Evdo96} 
N.V. Evdokimov, D.N. Klyshko, V.P. Komolov, and V.A.  Yarochkin, 
\newblock Phys. Usp. {\bf 39}, 83 (1996).

\bibitem{Cost09}
F.~Costa, N.~Harrigan, T.~Rudolph, and C.~Brukner,
\newblock arXiv.org:0902.0935v1  (2009).

\bibitem{Loud00}
R.~Loudon,
\newblock {\em The Quantum Theory of Light} (Oxford University Press, 2000).

\bibitem{Leon93}
U.~Leonhardt,
\newblock Phys. Rev. A {\bf 48}, 3265 (1993).

\bibitem{Aden03}
G.~Adenier and A.~Khrennikov,
\newblock arXiv:quant-ph/0306045  (2003).

\bibitem{Brun08}
N.~Brunner, C.~Branciard, and N.~Gisin,
\newblock Phys. Rev. A {\bf 78}, 052110 (2008).

\end{thebibliography}

\end{document}